\documentclass[twocolumn,showpacs,eqsecnum,amsmath,amssymb,pra]{revtex4}

\usepackage{graphicx}
\usepackage{dcolumn}
\usepackage{bm}

\begin{document}


\title {
Theoretical study of molecular electronic excitations and optical
transitions of C$_{60}$ }

\author {A. V. Nikolaev}
 \altaffiliation[Also at: ]{Institute of Physical
Chemistry of RAS, Leninskii pr. 31, 117915, Moscow, Russia. \\
e-mail: Alexander.Nikolaev@Algodign.com}

\author{I. V. Bodrenko}

\author{E. V. Tkalya}
\altaffiliation[Also at: ]{Institute of Nuclear Physics, Moscow
State University, 119992, Moscow, Vorob'evy Gory, Russia}

\affiliation{ Algodign LLC, Bolshaya Sadovaya 8/1, Moscow 123001,
Russia }

\pacs{31.25.-v, 36.20.Kd, 81.05.Tp}

\date{\today}

\begin{abstract}
We report results on {\it ab initio} calculations of excited
states of the fullerene molecule by using configuration
interaction (CI) approach with singly excited determinants (SCI).
We have used both the experimental geometry and the one optimized
by the density functional method and worked with basis sets at the
cc-pVTZ and aug-cc-pVTZ level. Contrary to the early SCI
semiempirical calculations, we find that two lowest $^1 T_{1u}
\leftarrow {}^1 A_g$ electron optical lines are situated at
relatively high energies of $\sim$5.8 eV (214 nm) and $\sim$6.3 eV
(197 nm). These two lines originate from two $^1 T_{1u} \leftarrow
{}^1 A_g$ transitions: from HOMO to (LUMO+1) ($6h_u \rightarrow
3t_{1g}$) and from (HOMO--1) to LUMO ($10h_g \rightarrow
7t_{1u}$). The lowest molecular excitation, which is the $1 ^3
T_{2g}$ level, is found at $\sim$2.5 eV. Inclusion of doubly
excited determinants (SDCI) leads only to minor corrections to
this picture. We discuss possible assignment of absorption bands
at energies smaller than $5.8$ eV (or $\lambda$ larger than
214~nm).
\end{abstract}

\pacs{31.25.-v, 36.20.Kd, 81.05.Tp}

\maketitle

\section {Introduction}
\label{sec:int}

The buckminsterfullerene molecule C$_{60}$ attracts much attention
of theoreticians since its discovery in 1985 \cite{Kro,Dre}.
Indeed, the role of the fullerene molecule in quantum chemistry is
unique because of its highest group ($I_h$) of symmetry leading to
a high degeneracy of molecular orbitals (MOs). The MOs of C$_{60}$
are classified according to irreducible representations (irreps)
of $I_h$, which are $a$, $t_1$, $t_2$, $g$, and $h$. (The
dimensions of the irreps are 1, 3, 3, 4, and 5, correspondingly.)
For example, the highest occupied molecular orbital (HOMO) is a
fivefold electron shell of $h_u$ symmetry while the lowest
unoccupied molecular orbital (LUMO) is a threefold level of
$t_{1u}$ symmetry.

High degeneracies of molecular orbitals imply a complex electronic
structure \cite{Pla,Jud,NM,WLT}. However, since all electron
shells of the neutral C$_{60}$ molecule are completely filled, the
ground state is of $^1 A_g$ symmetry and the many-electron states
of the molecule are not revealed. The situation changes when the
fullerene molecule is excited. Then a few electrons of C$_{60}$
are promoted to a higher energy, and some of electron shells
become open. The electrons of the open shells exhibit correlated
behavior and form many electron levels called molecular terms
\cite{Dre,Lar,Neg,Las,Bra}, which are also classified according to
the irreducible representations of the icosahedral symmetry ($A$,
$T_1$, $T_2$, $G$, and $H$). Thus, a careful study of these
correlations is needed for calculations of optical transitions
\cite{Lea,Gas} and molecular excitations. It is worth mentioning
that intramolecular correlations play an important role in alkali
doped fullerites, leading to superconductivity, magnetic behavior
and metal-insulator transitions \cite{Dre,Pra,For}.

Computations of many electron states of C$_{60}$ is a challenge
for the modern quantum theory \cite{Sza}. In the literature there
are few works devoted to the problem of calculations of molecular
excitations of the fullerene \cite{Lar,Neg,Las,Bra}, but to the
best of our knowledge all of them are semiempirical and are based
only on singly excited configuration interaction (CI). We
reproduce some of their computed levels in Table~\ref{tab0}. The
calculation of Negri {\it et al.}, Ref.~[\onlinecite{Neg}],
employed a semiempirical (QCFF/$\pi$) method, and took into
account 196 configuration state functions (CSF). The first excited
state (1$^3 T_{2g}$) was found at 2.06 eV, while 1$^1 T_{1u}$ and
2$^1 T_{1u}$ levels at 4.08 and 4.53 eV. A PPP-CI
(Pariser-Parr-Pople Configuration Interaction) calculation which
includes 134 CSF finds first two $^1 T_{1u}$ levels at 4.00 eV,
4.68 eV and 6.69 eV, with the first excited $^3 T_{2g}$ level at
2.23 eV. Finally, CNDO/S calculations of Braga {\it et al.} based
on 808 and 900 configurations predicts that 1$^1 T_{1u}$ is
lowered to 3.4 eV, while the three most intense transitions are at
4.38, 5.24 and 5.78 eV. Unfortunately, all of the CI calculations
reported in Refs.~[\onlinecite{Lar,Neg,Las,Bra}] are crucially
dependent on a number of approximations for matrix elements and
therefore cannot be considered as a true {\it ab initio} approach.
\begin{table}

\caption{ Selected molecular excitations calculated with
semiempirical CI calculations and in the present work. Energies
are in eV. $N_{CSF}$ stands for the number of configuration state
functions used for CI. \label{tab0} }
\begin{tabular}{c c c c c c}

\hline
Ref.& \multicolumn{2}{c}{ [\onlinecite{Las}] } & [\onlinecite{Neg}]& [\onlinecite{Bra}]& this work \\
method & \multicolumn{2}{c}{ PPP }   & QCFF/$\pi$ & CNDO/S &  {\it ab initio} SCI \\
$N_{CSF}$ & \multicolumn{2}{c}{ 134 }& 266        & 900    &  2857 \\
             &planar & 3D    &      &      &      \\
\hline
1$^3 T_{2g}$ & 2.456 & 2.232 & 2.06 &      & 2.549 \\
1$^1 G_u$    & 3.603 & 3.381 & 3.21 &      & 4.700 \\
1$^1 H_u$    & 3.635 & 3.391 & 3.15 &      & 4.771 \\
1$^1 T_{1u}$ & 4.227 & 4.000 & 4.08 & 3.40 & 5.796 \\
2$^1 T_{1u}$ & 4.750 & 4.681 & 4.53 & 4.06 & 6.335 \\

\hline

\end{tabular}
\end{table}

In the present study in comparison with
Refs.~[\onlinecite{Lar,Neg,Las,Bra}] we employ the {\it ab initio}
treatment, which represents a new level of CI calculations of the
fullerene molecule.

\section {Method of calculation}

\label{sec:met}

\subsection {Configuration Interaction}

A well known recipe for correlations is the method of
configuration interaction (CI) \cite{Sza}. The basic idea is to
diagonalize the $N$-electron Hamiltonian in a basis of
$N$-electron functions (Slater determinants). In principle, the
full CI method with infinite number of the molecular orbitals
provides an exact solution of the many-electron problem. In
practice, even for small molecules and moderately sized
one-electron basis sets, the number of $N$-electron determinants
is enormous. To avoid it, we first introduce an active space which
comprises a set of highest occupied molecular orbitals (HOMOs) and
a set of lowest unoccupied molecular orbitals (LUMOs). The space
of the MOs is divided into three subspaces: the inactive, the
active and the external orbitals. The inactive orbitals are double
occupied, while the external orbitals are unoccupied. Within the
chosen active space one can consider various approximations to the
complete active space CI matrix by truncating the many-electron
trial function at some excitation level. For example, singly
excited CI (SCI) approach has been proven to be adequate as a
first approximation for the neutral molecule because all electron
shells of C$_{60}$ are filled up. In the following for large
active spaces (AS) we use SCI, but also consider singly and doubly
excited determinants (SDCI) for smaller AS, and a complete active
space (CAS) CI method \cite{CAS} for important active spaces of
two molecular shells (8 MOs). The computer code for the CI
calculations is the modified and extended version of the earlier
program, Ref.~[\onlinecite{NM}].

\subsection {Computational details}

The CI calculations have been carried out with the set of
molecular orbitals (MOs) obtained from the restricted Hartree-Fock
(RHF) self-consistent-field calculation \cite{Sza} of the neutral
fullerene molecule. In the following we have used two sets of
coordinates of the C$_{60}$ molecule. First set was obtained in
Ref.~[\onlinecite{Gre}] (DFT/B3LYP method). Two different C-C bond
lengths were $R_5=1.4507$ {\AA} (C-C bond in pentagons) and
$R_6=1.3906$ {\AA} (short C-C bond in hexagons). The second set
corresponds to the experimental geometry \cite{expg} with
$R_5=1.448$ {\AA} and $R_6=1.404$~{\AA}.

We have used our original RHF computer program \cite{AlgoRHF},
which uses the resolution of the identity \cite{RI} (RI) method
for calculation of the two-electron Coulomb integrals. The series
of the RI (auxiliary) basis sets, HCNO$.x$, were designed for the
RI-convergent calculations the RHF total energies and the
electronic properties of molecules containing hydrogen, carbon,
nitrogen and oxygen elements as discussed in \cite{AlgoRHF}. Here
$x$ indicates (but not exactly equal to) the accuracy (in atomic
units) of the total RHF energy: smaller $x$ implies more RI
functions and more accurate RI basis set. The accuracy of the RI
basis set is almost independent of the choice of molecular basis
set. Having performed test calculations with Pople's 6-31G* MO
basis set \cite{Pop,baselib} and the HCNO$.x$ RI series we have
found out that a reasonable convergence is achieved for the
HCNO.001 RI basis set, which is used for all further calculations.

Throughout the paper we have adopted short notation
aug-cc-pVTZ(-f) for Dunning's augmented correlation consistent
polarizable valence triple zeta molecular basis set aug-cc-pVTZ
\cite{Dun,baselib} where all polarization functions of
$f$-symmetry are excluded from the set.
\begin{table}

\caption{ Electron shells of C$_{60}$. $E_{MO}$ is the
one-electron Hartree-Fock MO energy (in eV). The basis set is
aug-cc-pVTZ(-f). $E_{tot}=-61832.163$~a.u. \label{tab3} }

\begin{tabular}{l c c c r}

\hline
                 & MOs & symmetry & $E_{MO}$ \\
LUMO+18& 244-248 &    8$h_u$    &  4.303 \\
LUMO+17& 239-243 &   14$h_g$    &  3.855 \\
LUMO+16& 235-238 &    8$g_g$    &  3.794 \\
LUMO+15& 232-234 &   10$t_{2u}$ &  3.300 \\
LUMO+14& 229-231 &    4$t_{1g}$ &  3.280 \\
LUMO+13&  228    &    6$a_g$    &  3.161 \\
LUMO+12& 225-227 &    8$t_{1u}$ &  3.115 \\
LUMO+11& 220-224 &    7$h_u$    &  2.557 \\
LUMO+10& 215-219 &   13$h_g$    &  2.418 \\
LUMO+9 & 211-214 &    7$g_g$    &  2.254 \\
LUMO+8 & 206-210 &   12$h_g$    &  2.121 \\
LUMO+7 & 203-205 &    9$t_{2u}$ &  1.994 \\
LUMO+6 & 200-202 &    8$t_{2u}$ &  1.774 \\
LUMO+5 & 196-199 &    7$g_u$    &  1.751 \\
LUMO+4 &   195   &    5$a_g$    &  1.496 \\
LUMO+3 & 190-194 &   11$h_g$    &  1.398 \\
LUMO+2 & 187-189 &    7$t_{2u}$ &  1.150 \\
LUMO+1 & 184-186 &    3$t_{1g}$ &  0.947 \\
LUMO   & 181-183 &    7$t_{1u}$ & -0.817 \\

HOMO   & 176-180 &    6$h_u$    & -7.834 \\
HOMO-1 & 171-175 &   10$h_g$    & -9.627 \\
HOMO-2 & 167-170 &    6$g_g$    & -9.890 \\
HOMO-3 & 163-166 &    6$g_u$    &-12.589 \\
HOMO-4 & 160-162 &    6$t_{2u}$ &-13.050 \\
\hline

\end{tabular}
\end{table}

\section {Results and discussion}

\subsection {SCI excitation spectrum of C$_{60}$}

The main results are shown in Tables~\ref{tab3} and \ref{tab12}.
The (HOMO-4) to (LUMO+18) energies of C$_{60}$ calculated with the
aug-cc-pVTZ(-f) basis set are quoted in Table~\ref{tab3}. We then
define a broad active space spanned by these MOs and occupied with
42 correlated electrons and use it in our SCI calculations. Table
\ref{tab12} displays the resulting spectrum of lowest molecular
excitations. In order to evaluate the influence of two bond
lengths of the fullerene molecule, we have performed calculations
with two different geometries of C$_{60}$: optimized by DFT
\cite{Gre} and the experimental one \cite{expg}. Inspection of
Table~\ref{tab12} shows that the order of levels is mainly
conserved, while a typical numerical deviation is $\sim0.2$~eV.
From this we conclude that a possible small change of the C$_{60}$
bond lengths do not change appreciably its excitation spectrum.

The most striking feature of the calculation is the relatively
high energy position of the two lowest ${}^1 T_{1u}$ excitations
(given already in Table~\ref{tab0}) with oscillator strengths 1.02
and 0.16, respectively. For a review on oscillator strengths in
semiempirical calculations see Ref.~\cite{Wes}. In the following
we limit ourselves to the optimized geometry \cite{Gre} and
consider mainly the problem of convergency and adequacy of the SCI
calculation leaving discussion and conclusions until
Sec.~\ref{Dis}.
\begin{table}

\caption{ Lowest excitations of C$_{60}$. OG refers to the bond
lengths of C$_{60}$ optimized by DFT, Ref.~\onlinecite{Gre}, EG to
the experimental ones, Ref.~\onlinecite{expg}. \label{tab12} }

\begin{tabular}{c c c c c c c c c}

\hline

   & $^3 T_{2g}$ & $^3 T_{1g}$ & $^1 T_{2g}$ & $^3 H_g$ & $^1 T_{1g}$ & $^3 G_g$ & $^1 G_g$ & $^3 G_u$  \\
OG & 2.549       & 3.041       & 3.229       & 3.287    & 3.416       & 3.417    & 3.538    & 3.595 \\
EG & 2.384       & 2.877       & 3.051       & 3.100    & 3.239       & 3.235    & 3.357    & 3.464 \\
\hline

   & $^3 T_{1u}$ & $^3 T_{2u}$ & $^1 H_g$ & $^3 H_u$ & $^1 T_{2u}$ & $^1 G_u$ & $^1 H_u$ & $^3 T_{2u}$ \\
OG & 3.917       & 3.996       & 4.079    & 4.304    & 4.611       & 4.700    & 4.771    & 4.960    \\
EG & 3.771       & 3.801       & 3.906    & 4.154    & 4.447       & 4.534    & 4.588    & 4.822    \\
\hline

   & $^3 G_u$ & $^1 G_u$ & $^3 G_g$ & $^3 H_u$ & $^1 H_u$ & $^3 H_u$ & $^1 T_{1u}$ & $^3 T_{1u}$  \\
OG & 5.163    & 5.291    & 5.355    & 5.454    & 5.598    & 5.732    & 5.796       & 5.972      \\
EG & 5.023    & 5.157    & 5.300    & 5.295    & 5.445    & 5.662    & 5.667       & 5.849      \\
\hline

   & $^3 H_u$ & $^1 T_{2u}$ & $^3 G_u$ & $^3 T_{2g}$ & $^1 H_u$ & $^1 G_u$ & $^3 H_u$ & $^1 H_u$ \\
OG & 5.986    & 6.001       & 6.019    & 6.124       & 6.141    & 6.208    & 6.236    & 6.264 \\
EG & 5.823    & 5.872       & 5.873    & 6.041       & 5.954    & 6.021    & 6.188    & 6.218 \\
\hline

   & $^1 T_{1u}$ & $^3 G_u$ & $^3 T_{1u}$ & $^1 H_g$ & $^3 T_{2u}$ & $^1 T_{2g}$ & $^3 H_u$ & $^3 H_g$ \\
OG & 6.335       & 6.337    & 6.340       & 6.412    & 6.422       & 6.444       & 6.501    & 6.518 \\
EG & 6.182       & 6.216    & 6.206       & 6.340    & 6.275       & 6.360       & 6.395    & 6.400  \\

 \hline

\end{tabular}
\end{table}

\subsection {Basis set dependence}

To study basis set dependence of the SCI excitation spectrum we
narrowed active space to 167--219 MOs, see Table~\ref{tab3}. This
active space consists of 14 electron shells (from HOMO-2 to
LUMO+10) filled by 28 correlated electrons. Such calculations
require less computer time in comparison with the calculation with
the active space of Table~\ref{tab3} reported in
Table~\ref{tab12}, but lead to a less accurate molecular
excitation spectrum.

The results of such SCI calculations are presented in
Table~\ref{tab4}.
\begin{table}

\caption{ Energy spectrum of C$_{60}$ with various basis sets.
Basis sets are cc-pVDZ [DZ], aug-cc-pVDZ [aDZ)], cc-pVTZ [TZ)],
cc-pVTZ without the $f-$functions [TZ(-f)], and aug-cc-pVTZ(-f)
[aTZ(-f)], see text for details. \label{tab4} }

\begin{tabular}{c c c c c c c c c}

\hline
$\Gamma$     &6-31G* & DZ    & TZ    & TZ(-f)&$\Gamma$     & aDZ   &$\Gamma$     & aTZ(-f) \\

\hline
 $^3 T_{2g}$ & 2.593 & 2.575 & 2.582 & 2.573 & $^3 T_{2g}$ & 2.900 & $^3 T_{2g}$ & 2.669 \\
 $^3 T_{1g}$ & 3.114 & 3.071 & 3.053 & 3.048 & $^3 T_{1g}$ & 3.261 & $^3 T_{1g}$ & 3.107 \\
 $^1 T_{2g}$ & 3.322 & 3.276 & 3.249 & 3.248 & $^1 T_{2g}$ & 3.325 & $^1 T_{2g}$ & 3.254 \\
 $^3 H_g$    & 3.384 & 3.336 & 3.312 & 3.308 & $^3 H_g$    & 3.474 & $^3 H_g$    & 3.390 \\
 $^1 T_{1g}$ & 3.522 & 3.467 & 3.434 & 3.432 & $^1 T_{1g}$ & 3.541 & $^1 T_{1g}$ & 3.439 \\
 $^3 G_g$    & 3.540 & 3.481 & 3.447 & 3.445 & $^3 G_g$    & 3.547 & $^3 G_g$    & 3.477 \\
 $^1 G_g$    & 3.654 & 3.594 & 3.558 & 3.556 & $^1 G_g$    & 3.649 & $^1 G_g$    & 3.578 \\
 $^3 G_u$    & 3.672 & 3.675 & 3.700 & 3.685 & $^3 T_{1u}$ & 4.154 & $^3 G_u$    & 3.995 \\
 $^3 T_{1u}$ & 3.899 & 3.884 & 3.884 & 3.875 & $^3 T_{2u}$ & 4.196 & $^3 T_{1u}$ & 4.069 \\
 $^3 T_{2u}$ & 4.002 & 3.993 & 3.995 & 3.987 & $^1 H_g$    & 4.288 & $^3 T_{2u}$ & 4.132 \\
 $^1 H_g$    & 4.231 & 4.159 & 4.112 & 4.111 & $^3 G_u$    & 4.355 & $^1 H_g$    & 4.136 \\
 $^3 H_u$    & 4.383 & 4.365 & 4.365 & 4.356 & $^3 H_u$    & 4.754 & $^3 H_u$    & 4.541 \\
 $^1 T_{2u}$ & 4.650 & 4.632 & 4.625 & 4.618 & $^1 T_{2u}$ & 4.889 & $^1 T_{2u}$ & 4.751 \\

 $^1 G_u$    & 4.831 & 4.802 & 4.779 & 4.777 & $^1 G_u$    & 4.897 & $^1 G_u$    & 4.849 \\
 $^1 H_u$    & 4.862 & 4.835 & 4.816 & 4.811 & $^1 H_u$    & 4.976 & $^1 H_u$    & 4.899 \\
 $^3 T_{2u}$ & 5.120 & 5.075 & 5.053 & 5.045 & $^3 T_{2u}$ & 5.500 & $^3 T_{2u}$ & 5.241 \\
 $^3 G_u$    & 5.402 & 5.337 & 5.293 & 5.289 & $^3 G_u$    & 5.651 & $^3 G_u$    & 5.350 \\
 $^3 G_g$    & 5.443 & 5.398 & 5.392 & 5.373 & $^1 G_u$    & 5.771 & $^1 G_u$    & 5.451 \\
 $^1 G_u$    & 5.514 & 5.448 & 5.404 & 5.400 & $^3 H_u$    & 5.603 & $^3 H_u$    & 5.529 \\
 $^3 H_u$    & 5.576 & 5.524 & 5.490 & 5.486 & $^1 H_u$    & 5.722 & $^1 H_u$    & 5.652 \\
 $^1 H_u$    & 5.764 & 5.704 & 5.661 & 5.659 & $^1 T_{1u}$ & 6.116 & $^1 T_{1u}$ & 5.880 \\
 $^1 T_{1u}$ & 5.873 & 5.826 & 5.793 & 5.785 & $^3 H_u$    & 6.221 & $^3 G_g$    & 5.919 \\

 $^3 H_u$    & 5.940 & 5.859 & 5.810 & 5.797 & $^1 T_{2u}$ & 6.236 & $^1 T_{2u}$ & 6.071 \\
 $^3 T_{1u}$ & 6.057 & 5.992 & 5.946 & 5.937 & $^3 G_g$    & 6.249 & $^3 H_u$    & 6.075 \\
 $^3 H_u$    & 6.087 & 6.032 & 5.993 & 5.987 & $^3 T_{1u}$ & 6.280 & $^3 T_{1u}$ & 6.092 \\
 $^3 G_u$    & 6.116 & 6.067 & 6.030 & 6.022 & $^1 H_u$    & 6.254 & $^3 H_g$    & 6.098 \\
 $^1 T_{2u}$ & 6.165 & 6.096 & 6.048 & 6.046 & $^1 G_u$    & 6.281 & $^3 G_u$    & 6.103 \\
 $^3 T_{2g}$ & 6.194 & 6.113 & 6.054 & 6.042 & $^3 T_{2g}$ & 6.543 & $^1 H_u$    & 6.198 \\
 $^1 H_u$    & 6.299 & 6.246 & 6.202 & 6.199 & $^1 T_{1u}$ & 6.577 & $^1 G_u$    & 6.264 \\
 $^1 G_u$    & 6.363 & 6.314 & 6.274 & 6.271 & $^3 G_u$    & 6.954 & $^3 T_{2g}$ & 6.325 \\
 $^3 G_u$    & 6.445 & 6.386 & 6.346 & 6.340 & $^3 H_g$    & 6.960 & $^1 T_{1u}$ & 6.371 \\
 $^3 T_{1u}$ & 6.457 & 6.402 & 6.360 & 6.356 & $^3 T_{1u}$ & 7.118 & $^3 T_{1u}$ & 6.413 \\
 $^1 T_{1u}$ & 6.462 & 6.407 & 6.365 & 6.362 & $^1 T_{1u}$ & 7.122 & $^3 G_u$    & 6.452 \\
\hline

\end{tabular}
\end{table}
Notice that the order of the levels is almost the same for the
group of 6-31G*, cc-pVDZ, cc-pVTZ and cc-pVTZ(-f) basis sets. The
deviations of the molecular excitation energies are rather small,
of the order of 0.1~eV. The molecular energies are close to a
convergence and the basis f-functions have a small effect on
calculated values. On the other hand the inclusion of diffuse
functions of aug-cc-pVDZ(-f) and aug-cc-pVTZ(-f) leads to a
rearrangement of electronic levels, Table~\ref{tab4}, which
indicates that aug-cc-pVTZ(-f) is the best set for SCI
calculations.

\subsection {Active space dependence}

Here we study the dependence of SCI excitation energies on the
choice of active space. In Table~\ref{tab6} we keep the lower
boundary active shell, which is the HOMO-4 (6$t_{2u}$) level
unchanged and systematically increase the upper electron boundary.
Inspection of Table \ref{tab6} shows that we are close to a
convergence for the broad active space of 160--248 MOs, shown in
Table~\ref{tab3}. Therefore, we do not expect drastic changes if
the active space is increased even further. Notice however, that
the optically active $1 {}^1 T_{1u}$ and $2 {}^1 T_{1u}$ levels
monotonically decrease with the increase of active space.
\begin{table}

\caption{ Dependence of selected energy levels on chosen active
space. The active space consists of the MOs between lower and
upper MOs. The lower MO is 160, which corresponds to the HOMO-4
($6t_{2u}$) level. The upper MO changes from LUMO+1 to LUMO+18.
The basis set is aug-cc-pVTZ($-f$). \label{tab6} }

\begin{tabular}{c c c c c c c c c }

\hline
 upper MO   & 186   & 189   & 194   & 199   & 202   & 205   & 210   & 214   \\

$1^3 T_{2g}$ & 2.889 & 2.889 & 2.889 & 2.889 & 2.887 & 2.833 & 2.713 & 2.712 \\
$1^1 G_u$    & 4.808 & 4.808 & 4.808 & 4.808 & 4.807 & 4.801 & 4.758 & 4.758 \\
$1^1 H_u$    & 4.898 & 4.898 & 4.898 & 4.898 & 4.898 & 4.891 & 4.836 & 4.836 \\
$1^1 T_{1u}$ & 6.230 & 6.230 & 6.230 & 6.230 & 6.230 & 6.212 & 6.028 & 6.026 \\
$2^1 T_{1u}$ & 7.210 & 7.210 & 7.194 & 7.172 & 7.112 & 6.936 & 6.421 & 6.419 \\
\hline
 upper MO   & 219   & 224   & 227   & 231   & 234   & 238   & 243   & 248   \\

$1^3 T_{2g}$ & 2.633 & 2.633 & 2.623 & 2.618 & 2.600 & 2.599 & 2.584 & 2.549 \\
$1^1 G_u$    & 4.730 & 4.730 & 4.727 & 4.721 & 4.719 & 4.719 & 4.715 & 4.700 \\
$1^1 H_u$    & 4.802 & 4.802 & 4.798 & 4.793 & 4.791 & 4.790 & 4.785 & 4.771 \\
$1^1 T_{1u}$ & 5.862 & 5.862 & 5.862 & 5.860 & 5.839 & 5.838 & 5.816 & 5.796 \\
$2^1 T_{1u}$ & 6.354 & 6.354 & 6.352 & 6.352 & 6.348 & 6.348 & 6.343 & 6.335 \\

\hline

\end{tabular}
\end{table}

\subsection {Beyond SCI approximation for small AS}

\label{CAS}

We recall that all the results reported earlier have been obtained
under approximation that only singly excited configurations (SCI)
are taken into account. It is therefore instructive to estimate
the accuracy of this assumption. Since the active space shown in
Table~\ref{tab3} is very big, we study this problem by considering
fewer active MOs. We limit ourselves to four most important active
spaces: HOMO+LUMO ($6h_u+7t_{1u}$), HOMO+(LUMO+1)
($6h_u+3t_{1g}$), (HOMO--1)+LUMO ($10h_g+7t_{1u}$), and
(HOMO--1)+(LUMO+1)($10h_g+3t_{1g}$). All of them consists of 8
active MOs (five from $h$ and three from $t_1$), which accommodate
10 correlated electrons. For each of the active spaces we have
performed four calculations: SCI, singly and double excited CI
(SD-CI), singly, doubly and triply excited CI (SDT-CI), and with
all possible excitations. The latter is so called complete active
space (CAS-CI) calculation \cite{CAS}. In Tables
\ref{tab8}-\ref{tab11} we give the results for the four cases,
respectively. Notice that the CAS-CI results are very close to
that of SDT-CI, while both SCI and SDCI molecular energies miss
the precise values by a typical error of $\sim0.1$~eV.
Furthermore, as a rule SDCI energies overestimate the real values.
It is very probable that the same conclusions are applicable to a
calculation with a large AS. Indeed, our SDCI calculations
\cite{CI_sym} with a large AS of 186-248 MOs unambiguously
indicate an increase of ${}^1T_{1u}$ energies in comparison with
the SCI values: the energy of the first transition changes from
5.997 to 6.291 eV, while the energy of the second from 6.395 to
7.154 eV. Thus the SDCI results underline the importance of SDT-CI
calculation for the C$_{60}$ molecule.
\begin{table}

\caption{ Lowest excitations of C$_{60}$, calculated with 10
correlated electrons in the $6h_u + 7t_{1u}$ limited active space
(see text for details). \label{tab8} }

\begin{tabular}{c c c c c c c c}

\hline
excitation level    & S     & SD    & SDT    & all (CAS) \\
$^1 A_g$    & 0     & 0     & 0      & 0       \\
$^3 T_{2g}$ & 3.312 & 3.383 & 3.338 & 3.338 \\
$^3 T_{1g}$ & 3.482 & 3.551 & 3.511 & 3.511 \\
$^1 T_{2g}$ & 3.590 & 3.661 & 3.622 & 3.623 \\
$^3 H_g$    & 3.640 & 3.699 & 3.651 & 3.651 \\
$^1 T_{1g}$ & 3.691 & 3.761 & 3.713 & 3.713 \\
$^3 G_g$    & 3.709 & 3.768 & 3.726 & 3.726 \\
$^1 G_g$    & 3.769 & 3.833 & 3.791 & 3.791 \\
$^1 H_g$    & 4.298 & 4.337 & 4.281 & 4.280 \\
\hline

\end{tabular}
\end{table}
\begin{table}

\caption{ Lowest excitations of C$_{60}$, calculated with 10
correlated electrons in the $6h_u + 3t_{1g}$ limited active space
(see text for details). \label{tab9} }

\begin{tabular}{c c c c c c c c}

\hline
excitation level    & S     & SD    & SDT    & all (CAS)     \\
$^1 A_g$    & 0     & 0     & 0      & 0      \\
$^3 T_{2u}$ & 5.188 & 5.322 & 5.239 & 5.239 \\
$^3 T_{1u}$ & 5.234 & 5.368 & 5.297 & 5.298 \\
$^3 G_u$    & 5.410 & 5.544 & 5.470 & 5.471 \\
$^3 H_u$    & 5.461 & 5.595 & 5.529 & 5.530 \\
$^1 H_u$    & 5.550 & 5.684 & 5.616 & 5.617 \\
$^1 G_u$    & 5.594 & 5.728 & 5.661 & 5.662 \\
$^1 T_{2u}$ & 5.692 & 5.826 & 5.760 & 5.761 \\
$^1 T_{1u}$ & 6.443 & 6.577 & 6.438 & 6.439 \\
\hline

\end{tabular}
\end{table}

\begin{table}

\caption{ Lowest excitations of C$_{60}$, calculated with 10
correlated electrons in the $10h_g + 7t_{1u}$ limited active space
(see text for details). \label{tab10} }

\begin{tabular}{c c c c c c c c}

\hline
excitation level    & S     & SD    & SDT    & all (CAS) \\
$^1 A_g$    & 0     & 0     & 0      & 0       \\
$^3 G_u$    & 5.218 & 5.497 & 5.331 & 5.334 \\
$^3 T_{1u}$ & 5.233 & 5.512 & 5.371 & 5.374 \\
$^3 T_{2u}$ & 5.440 & 5.719 & 5.562 & 5.565 \\
$^3 H_u$    & 5.457 & 5.737 & 5.596 & 5.600 \\
$^1 T_{2u}$ & 5.548 & 5.828 & 5.679 & 5.682 \\
$^1 G_u$    & 5.580 & 5.860 & 5.716 & 5.720 \\
$^1 H_u$    & 5.605 & 5.884 & 5.737 & 5.741 \\
$^1 T_{1u}$ & 6.997 & 7.277 & 6.959 & 6.960 \\
\hline

\end{tabular}
\end{table}
\begin{table}

\caption{ Lowest excitations of C$_{60}$, calculated with 10
correlated electrons in the $10h_g + 3t_{1g}$ limited active space
(see text for details). \label{tab11} }

\begin{tabular}{c c c c c c c c}

\hline
excitation level    & S     & SD    & SDT    & all (CAS) \\
$^1 A_g$    & 0     & 0     & 0      & 0       \\
$^3 H_g$    & 7.072 & 7.110 & 7.075 & 7.075 \\
$^3 T_{2g}$ & 7.116 & 7.154 & 7.125 & 7.125 \\
$^3 G_g$    & 7.174 & 7.224 & 7.194 & 7.194 \\
$^3 T_{1g}$ & 7.192 & 7.236 & 7.209 & 7.209 \\
$^1 T_{2g}$ & 7.299 & 7.348 & 7.320 & 7.320 \\
$^1 T_{1g}$ & 7.337 & 7.383 & 7.353 & 7.353 \\
$^1 G_g$    & 7.388 & 7.442 & 7.415 & 7.415 \\
$^1 H_g$    & 7.868 & 7.898 & 7.859 & 7.859 \\
\hline

\end{tabular}
\end{table}

It is worth noting that two lowest $^1 T_{1u} \leftarrow {}^1A_g$
transitions can be found already in Tables \ref{tab9} and
\ref{tab10}. Their energies are quite high, 6.44~eV ($6h_u
\rightarrow 3t_{1g}$) and 6.96~eV ($10h_g \rightarrow 7t_{1u}$).
If both of the $^1T_{1g}$ levels are included in a CI calculation,
they interact with each other, because they are of the same
symmetry. Nevertheless, the lowest $1 ^1T_{1u}$ level stays
predominantly of the $6h_u \rightarrow 3t_{1g}$ origin, while the
$2 ^1T_{1u}$ level is mainly of the $10h_g \rightarrow 7t_{1u}$
character. Notice that the lowest part of molecular excitations of
C$_{60}$ stems from the HOMO+LUMO ($6h_u+7t_{1u}$) active space,
Table~\ref{tab8}.

\section {Discussion and Conclusions}

\label{Dis}

We have found that the lowest electron optical transitions are at
$\sim5.8$~eV (214 nm) and $\sim 6.3$~eV (197 nm), which are at
disagreement with the results obtained with the early
semiempirical CI calculations \cite{Neg,Las,Bra}. We have analyzed
the influence on the SCI excitation spectrum from the
approximations used in our calculation and concluded that all of
them are unlikely to change the computed values substantially.
Thus, according to our calculations, the first allowed optical
transition of C$_{60}$ lies very close to the characteristic
interstellar absorption at 217 nm \cite{Kro,Rab}.
%
\begin{figure}
\vspace{1mm} \resizebox{0.50\textwidth}{!} {
 \includegraphics{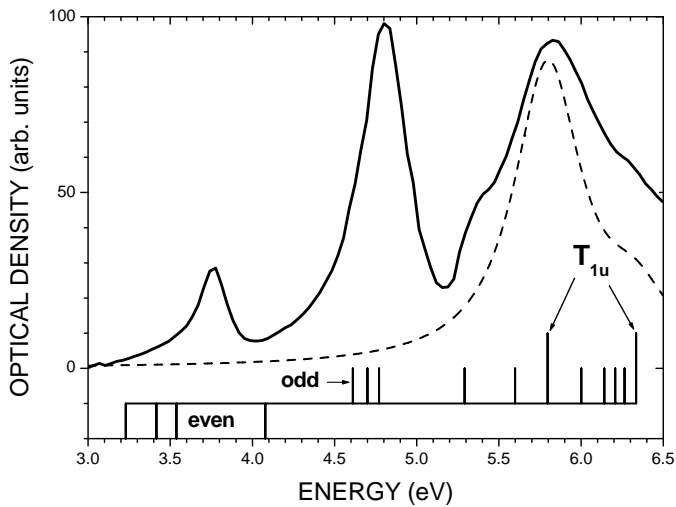}
}
\vspace{1mm} \caption{ Comparison between the experiment data on
optical density, Ref.~\onlinecite{Lea}, and the calculated spin
singlet lines of C$_{60}$. The dashed line is the contribution
from two $T_{1u}$ transitions broadened with the line width of
0.5~eV. The MO basis set is aug-cc-pVTZ(-f), the bond lengths are
taken from Ref.~\onlinecite{Gre}. } \label{fig1}
\end{figure}
%

From the {\it ab initio} study we come to the conclusion that
early semiempirical configuration interactions of the fullerene
molecule, Refs.~[\onlinecite{Neg,Las,Bra}] result in
systematically smaller excitation energies, see Table~\ref{tab0}.
This was also the case for the molecular ion C$_{60}^{2-}$, for
which the ${}^1 A_g$ spin singlet ground state was obtained
\cite{Neg2}. As shown in Refs.~[\onlinecite{NM,WLT}] the ground
state is rather the $^3 T_{1g}$ spin triplet in accordance with
Hund's rules.

Calculated oscillator strengths of two lowest optical transitions
to $1 {}^1 T_{1u}$ and $2 {}^1 T_{1u}$ states in the largest AS
(Table~\ref{tab3}) are 1.02 and 0.16, respectively. QCFF/$\pi$
method employed by Negri {\it et al.} \cite{Neg} resulted in 0.61
and 0.41 for oscillator strengths of two lowest transitions while
Braga {\it et al.} \cite{Bra} [CNDO/S method] obtained 0.08 and
0.41. As we remarked before our {\it ab initio} energies of two
lowest ${}^1 T_{1u}$ transitions are very different from the
semiemperical ones and a direct comparison of the oscillator
strengths is probably premature. On the other hand, early
semiemperical calculations give only a vague correspondence with
the experimental optical density \cite{Lea}, see Fig.~14 of
Ref.~\cite{Wes}. In order to achieve a good comparison Westin {\it
et al.} used a screened response oscillator strength distribution
calculated with an adjustable parameter $\nu$ \cite{Wes}. Under
such conditions the theoretical curve reproduces the three
experimental bands shown in Fig.~1. Our present {\it ab initio}
study accounts only for the 6~eV band, Fig.~1. Thus, two other
bands are probably of the electron-vibronic origin \cite{Her}. The
situation is not uncommon for complex molecules and it occurs for
example at the ${}^1 B_{2u} \leftarrow {}^1A_{1g}$ (3200 {\AA})
transition of naphthalene as discussed in Chapter II$\gamma$ of
Ref.~\cite{Her}.

The optical spectrum of C$_{60}$ also shows some similarities with
that of benzene \cite{Her,Hash}. Lowest excited states of benzene
are spin triplets, while the first electronically allowed optical
transition ($1 {}^1 E_{1u} \leftarrow {}^1 A_{1g}$) is situated at
relatively high energy of $\sim$7~eV \cite{Hash,Hir}. In benzene
two lowest absorption bands at 4.90 and 6.20 eV were unambiguously
assigned to the excited states of $1 {^1}B_{2u}$ and $1
{^1}B_{1u}$, respectively \cite{Hash,Hir}. These bands are
electronically forbidden and occur only due to the
electron-vibronic coupling, which underlines the importance of the
Herzberg-Teller mechanism \cite{Her} for the molecule. It is
plausible that the same scenario applies to the C$_{60}$
fullerene, and optical lines at smaller energies are caused by
vibronic coupling to numerous electron states in these energy
regions, Fig.~1. Indeed, there are indications that optical bands
observed in the low region of the absorption spectrum of C$_{60}$
can be interpreted as false origins of states of the $T_{1g}$
symmetry \cite{Neg3}. It is worth noting that singlet and triplet
states of C$_{60}$ can be distinguished by low energy electron
spectroscopy \cite{Doe}.

Unfortunately, the electronic spectrum of C$_{60}$ is not so well
understood as benzene's, which has a long history of calculations
and interpretations \cite{Her,Hash}. Although in all our SCI and
SD-CI calculations the $1 ^1 T_{1u}$ level remains at high energy
of $\sim$6 eV, we think that there is still a possibility for it
to be lowered if a calculation with a higher level of excitations
within the CI approach is performed (at least SDT-CI).
Interestingly, this is the case for the first allowed optical
transition ($1 {}^1 E_{1u} \leftarrow {}^1 A_{1g}$) in benzene.
The CASSCF calculation of C$_6$H$_6$ (6 $\pi$ electrons
distributed among 12 $\pi$ MOs) places the $1 {}^1 E_{1u}$ level
at a high energy of 8.77 eV, while a multireference
M{\o}ller-Plesset perturbation theory (MRMP) on the CASSCF states
lowers this energy by 1.84~eV to 6.93~eV (Table~2 of
Ref.~[\onlinecite{Hash}]). The last value is in perfect agreement
with the experiment (6.94~eV, Ref.~[\onlinecite{Hir}]). Notice
that MRMP is known for effectively accounting for high
excitations.

In conclusion, our S-CI and SD-CI calculations indicate that the
first electronically allowed transition $1 {}^1 T_{1u} \leftarrow
{}^1 A_{1g}$ is located at a relatively high energy of 5.8-6.0 eV.
Our finding opens up a very interesting question on assignment of
two lowest absorption bands at 3.8 and 4.8 eV found experimentally
\cite{Lea,Gas}. There are two ways to reconcile our calculations
with the experiment. First, the $1 {}^1 T_{1u}$ excited state can
be lowered at the level of SDT-CI or more refined CI calculations.
The second scenario is the electron-vibronic coupling
(Herzberg-Teller mechanism) \cite{Her}, which is the case for
benzene. Further investigations and {\it ab initio} calculations
are needed to clarify this issue.





\end{document}